\documentclass[journal=nalefd,manuscript=letter, layout=twocolumn]{achemso}                 
\usepackage{natmove}
\usepackage{graphicx,color,xcolor}
\usepackage{caption}
\usepackage{float}
\usepackage{geometry}
\usepackage{natbib}
\usepackage{setspace}
\usepackage{xkeyval}
\usepackage{array}
\usepackage{booktabs}
\usepackage{listingsutf8}
\usepackage{lmodern}
\usepackage{mathpazo}
\usepackage{microtype}
\usepackage[colorlinks=true,bookmarks=false,linkcolor=blue,citecolor=blue]{hyperref}
\let\oldmaketitle\maketitle
\let\maketitle\relax

\title{Gate-tunable two-dimensional superlattices in graphene}
\date{\today}

\author{Robin Huber}
\affiliation{Institute of Experimental and Applied Physics, University of Regensburg, D-93040 Regensburg, Germany}

\author{Ming-Hao Liu}
\email{minghao.liu@phys.ncku.edu.tw}
\affiliation{Department of Physics, National Cheng Kung University, Tainan 70101, Taiwan}

\author{Szu-Chao Chen}
\affiliation{Department of Physics, National Cheng Kung University, Tainan 70101, Taiwan}

\author{Martin Drienovsky}
\affiliation{Institute of Experimental and Applied Physics, University of Regensburg, D-93040 Regensburg, Germany}

\author{Andreas Sandner}
\affiliation{Institute of Experimental and Applied Physics, University of Regensburg, D-93040 Regensburg, Germany}

\author{Kenji Watanabe}
\affiliation{National Institute for Materials Science, 1-1 Namiki, Tsukuba 305-0044, Japan}

\author{Takashi Taniguchi}
\affiliation{National Institute for Materials Science, 1-1 Namiki, Tsukuba 305-0044, Japan}

\author{Klaus Richter}
\affiliation{Institute of Theoretical Physics, University of Regensburg, D-93040 Regensburg, Germany}

\author{Dieter Weiss}
\affiliation{Institute of Experimental and Applied Physics, University of Regensburg, D-93040 Regensburg, Germany}

\author{Jonathan Eroms}
\email{jonathan.eroms@ur.de}
\affiliation{Institute of Experimental and Applied Physics, University of Regensburg, D-93040 Regensburg, Germany}

\begin{document}

\begin{tocentry}
	\includegraphics[width=1\linewidth]{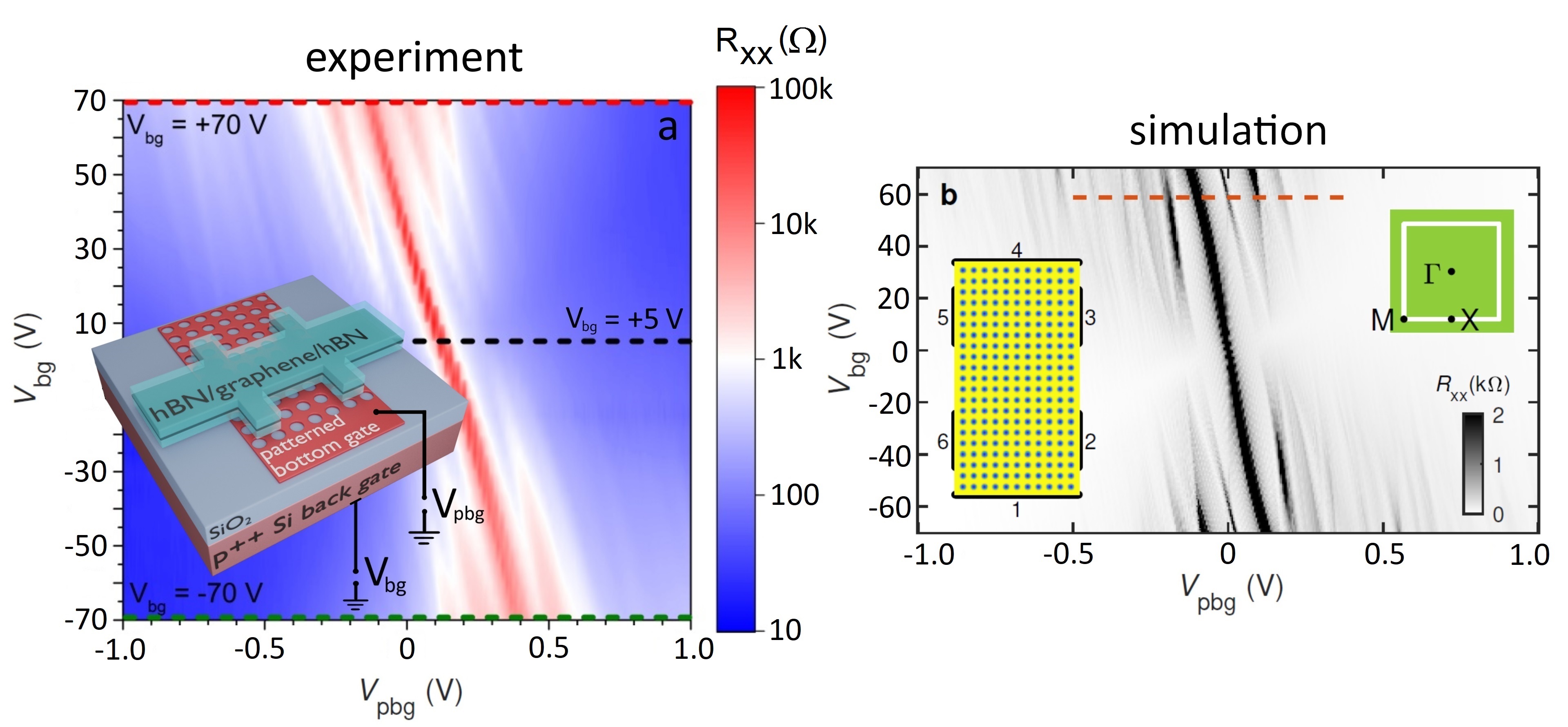}
\end{tocentry}

\twocolumn[
\begin{@twocolumnfalse}
	\oldmaketitle
	\begin{abstract}
		We report an efficient technique to induce gate-tunable two-dimensional superlattices in graphene by the combined action of a back gate and a few-layer graphene patterned bottom gate complementary to existing methods. The patterned gates in our approach can be easily fabricated and implemented in van der Waals stacking procedures allowing flexible use of superlattices with arbitrary geometry. In transport measurements on a superlattice with lattice constant $a=40$ nm well pronounced satellite Dirac points and signatures of the Hofstadter butterfly including a non-monotonic quantum Hall response are observed. Furthermore, the experimental results are accurately reproduced in transport simulations and show good agreement with features in the calculated band structure. Overall, we present a comprehensive picture of graphene-based superlattices, featuring a broad range of miniband effects, both in experiment and in theoretical modeling. The presented technique is suitable for studying more advanced geometries which are not accessible by other methods.
	\end{abstract}
\end{@twocolumnfalse}
]

\maketitle
\normalfont

Electrons in an artificially created periodic potential---a superlattice---follow Bloch's theorem, and they reside in minibands\cite{Keldysh1963,EsakiTsu1970}. Artificial superlattices are particularly suitable to study effects inaccessible in natural crystals, such as Hofstadter's butterfly - a fractal band structure electrons exhibit by adding a uniform, perpendicular magnetic field and which is governed by the ratio of magnetic flux $\phi$ per lattice unit cell and the magnetic flux quantum $\phi_0 = h/e$. For strong lattice potential and weak magnetic field, Hofstadter first calculated the band structure, which shows continuous bands for rational $\phi/\phi_0 = p/q$ with $p$ and $q$ co-prime integers and exhibit a self-similar structure, which was coined Hofstadter butterfly due to its peculiar shape.\cite{Hofstadter1976} For the opposite case of strong magnetic field and weak lattice potential, the band structure is controlled by the inverse magnetic flux $\phi_0/\phi = q/p$ instead \cite{MacDonald1983}. 
As was first shown by Thouless {\em et al.} the transverse resistance is quantized, whenever the Fermi level lies within a gap, either in between Landau levels, or within the Hofstadter spectrum \cite{Thouless1982}. This observation formed the theoretical basis for understanding the precise quantization of the quantum Hall plateaus, using notions of topology\cite{Niu1985,Aoki1986}. For a system where the Hofstadter spectrum is present, the filling factor $\nu$ and quantum Hall plateau values are linked by a Diophantine equation, \cite{Wannier1978,Thouless1982,MacDonald1983}
\begin{equation}
\nu=(\phi_{0}/\phi)s+t,
\label{Eq:Dio}
\end{equation}
where $s$ and $t$ are integers such that $R_{xy}=\frac{h}{t\,e^2}$. Note that $s=0$ represents the standard quantum Hall physics ($\nu=t$),  and values $s\ne 0$ are associated with internal gaps of the Hofstadter butterfly. In the latter case, the transverse quantum Hall resistance does not follow the filling factor $\nu$ in a monotonic way \cite{Thouless1982}. Since in the recursive Hofstadter spectrum there are infinitely many gaps, the transverse resistance will show more and more non-monotonic steps, as an increasing number of gaps is resolved in experiment.

\begin{figure*}
\includegraphics[width=1\linewidth]{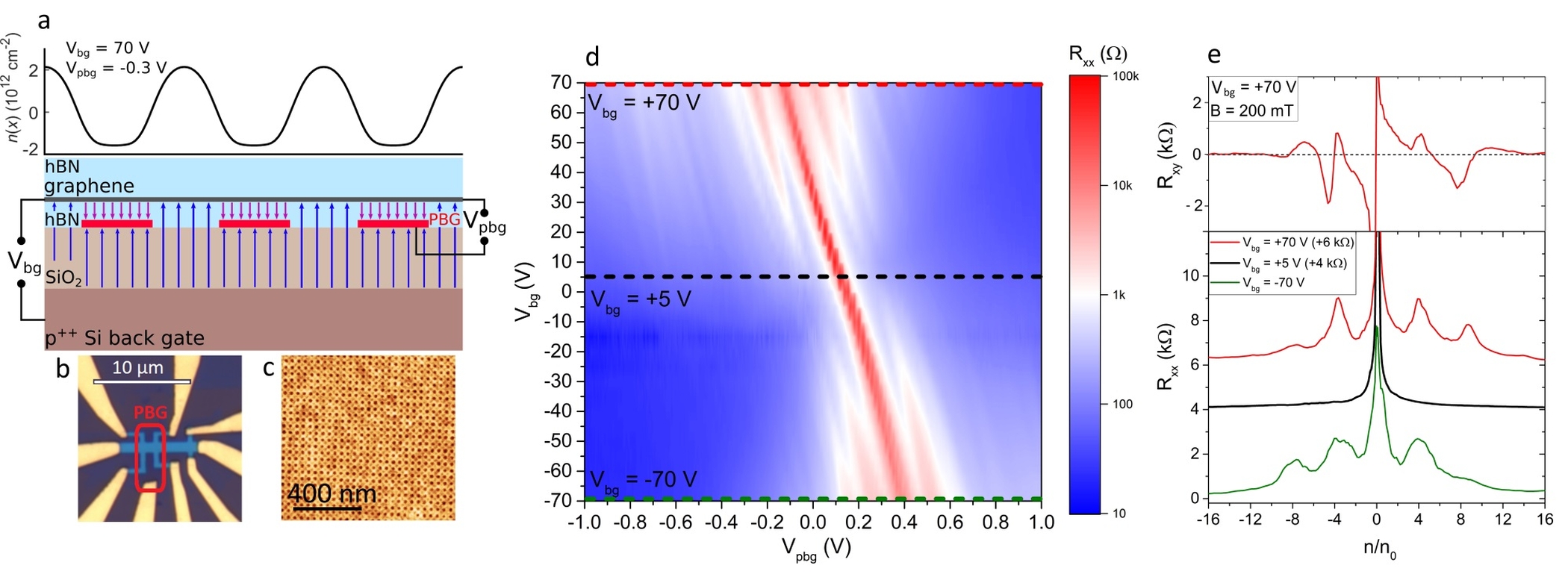}
\caption{\textbf{Sample Layout and Gate Response at Zero Magnetic Field.} \textbf{a} Schematic of the sample geometry. Due to the interplay between a Si back gate and a few-layer graphene patterned bottom gate (PBG), a periodic charge carrier density modulation can be induced in the encapsulated graphene layer on top of the two gates. \textbf{b} Micrograph of the studied device in Hall bar geometry and Cr/Au edge contacts. Red line marking the position of the PBG. \textbf{c} AFM picture of the PBG with a square superlattice and a lattice constant of $a=40$ nm. For the studied sample, a periodic array of holes was etched into a bilayer graphene flake. \textbf{d} Gate map of the device at $T=1.5$ K. Longitudinal resistance $R_{xx}$ as a function of back gate voltage $V_\mathrm{bg}$ and PBG voltage $V_\mathrm{pbg}$. By increasing the back gate voltage, {\em i.e.}, increasing the modulation strength, satellite peaks start to occur besides the main Dirac peak. \textbf{e} Linecuts at three different back gate voltages highlight the additional features upon tuning the periodic potential strength. Upper panel shows the sign change of the corresponding Hall resistance $R_{xy}$ at the position of the satellite Dirac points at $B=200$ mT.}
\label{fig:ExpZero}
\end{figure*}

First efforts to detect the Hofstadter band structure were performed in high-mobility GaAs-based two-dimensional (2D) electron gases \cite{Albrecht2001,Geisler2004}, showing minigaps, and also non-monotonic Hall curves. However, the lattice period was limited to $a\ge100$ nm due to technological constraints. Utilizing moir\'{e} lattices formed in graphene/hBN heterostructures\cite{Dean2010,Decker2011,Yankowitz2012,Wang1Dcontacts2013}, the Hofstadter butterfly was unambiguously demonstrated\cite{Dean2013,Hunt2013,Ponomarenko2013}. However, the maximum lattice period in this case is limited to about 14 nm, given by the lattice mismatch between hBN and graphene. This means that the condition of one magnetic flux quantum per unit cell area requires at least 25 T, out of scope for standard laboratory magnets.  In addition, the lattice symmetry is determined to be hexagonal, and the potential is not gate-tunable. Finally, even though a complete theoretical description of moir{\'e} lattices was presented\cite{Wallbank2013,Moon2014}, many terms including strain or local gaps have to be included, which are poorly controlled experimentally. 
2D superlattices in graphene were also realized in etched graphene antidot lattices\cite{EromsNJP2009}, showing ballistic transport\cite{Sandner2015,Yagi2015,Power2017,Datseris2019} and superlattice Dirac points\cite{ParkPeriodic2008,Jessen2019}, but due to the strong potential, which cannot be tuned, more subtle effects are not visible. Therefore, a more flexible method with a gate-defined superlattice potential is required. A first approach was recently pursued by Forsythe {\em et al.}, based on a patterned dielectric\cite{Forsythe2018} and demonstrating gate-tunable superlattice effects. Here, both square and hexagonal lattices were realised, resulting in satellite Dirac peaks\cite{ParkPeriodic2008} and observation of gaps in the Hofstadter butterfly. However, the characteristic non-monotonic transverse resistance in the high magnetic field regime, predicted by Thouless {\em et al.} \cite{Thouless1982}, was not reported, requiring further improvements in device quality. \cite{ForsythePhD}

In this work, combining uniform and patterned gates,\cite{DrienovskyPBG2017,Drienovsky2018} we demonstrate  satellite resistance peaks corresponding to Dirac cones to fourth order, and the Hofstadter butterfly, including a non-monotonic quantum Hall response. Our samples (see Fig.~\ref{fig:ExpZero}a) consist of a single-layer graphene sheet encapsulated in hBN and placed on top of a few-layer graphene patterned bottom gate (PBG) residing on an oxidized silicon substrate. Importantly, the lower hBN is only 5 nm thin to prevent damping of the electrostatic potential. In addition, the doped Si serves as a uniform back gate. In principle, patterned graphite gates can be easily fabricated by standard electron beam lithography and oxygen plasma etching and can be integrated in common van der Waals stacking and transfer processes which allows great flexibility in fabricating devices including picking up and transferring pre patterned graphite gates. More details are given in the Supporting Information and previous reports\cite{DrienovskyPBG2017,Drienovsky2018}. By using this gating scheme, a widely tunable 2D periodic charge carrier density modulation can be induced in the graphene layer ranging from unipolar to bipolar potential profile with tunable potential strength (see Supporting Information for gate maps of polarity regime, modulation strength and average carrier density and exemplary potential profiles). The exact potential shape in the system can be determined from elementary electrostatics\cite{MingHaoCap2013}, allowing for a detailed comparison between miniband structure and calculated transport characteristics.

Figure~\ref{fig:ExpZero}b shows a  micrograph of the studied device with red outlining the position of the PBG, and in Fig.~\ref{fig:ExpZero}c, an atomic force microscopy (AFM) image of the PBG with a square lattice of period $a=40$ nm can be seen. Figure~\ref{fig:ExpZero}d shows the gate response of the longitudinal resistivity $R_{xx}$ at a magnetic field of $B=0$ T and a temperature of $T=1.5$ K. As can clearly be seen, the PBG voltage $V_\mathrm{pbg}$ largely controls the overall carrier density, and the Si back gate voltage $V_\mathrm{bg}$ mainly sets the modulation strength. For $V_\mathrm{bg} \approx 0$ V, only one Dirac peak is visible, as expected for intrinsic graphene. The field effect mobility extracted at low back gate voltage is about 40\,000 cm\textsuperscript{2}/Vs. By increasing $V_\mathrm{bg}$, {\em i.e.,} increasing the potential strength, several pronounced satellite peaks start to appear next to the main Dirac peak. Figure~\ref{fig:ExpZero}e shows $R_{xx}$ plotted as a function of normalised charge carrier density $n/n_{0}$ (with $n_{0}=1/a^2$) which gives the number of electron/holes per superlattice unit cell area for three different back gate voltages $V_\mathrm{bg}=5$ V and $V_\mathrm{bg}=\pm70$ V. The most pronounced satellite peaks manifest when one superlattice unit cell is filled up with 4 and 8 electrons/holes which reflects the 4-fold degeneracy of graphene. These satellite peaks are accompanied by sign changes of the Hall resistance $R_{xy}$ at a small magnetic field of $B=200$ mT (see upper panel of Fig.~\ref{fig:ExpZero}e). This confirms that the carrier type changes between electron and hole conduction, and proves the existence of well-defined superlattice-induced minibands. Also, when the sign of the back gate voltage is reversed, the satellite Dirac peaks are mirrored with respect to the main Dirac peak.

\begin{figure*}
\includegraphics[width=1\linewidth]{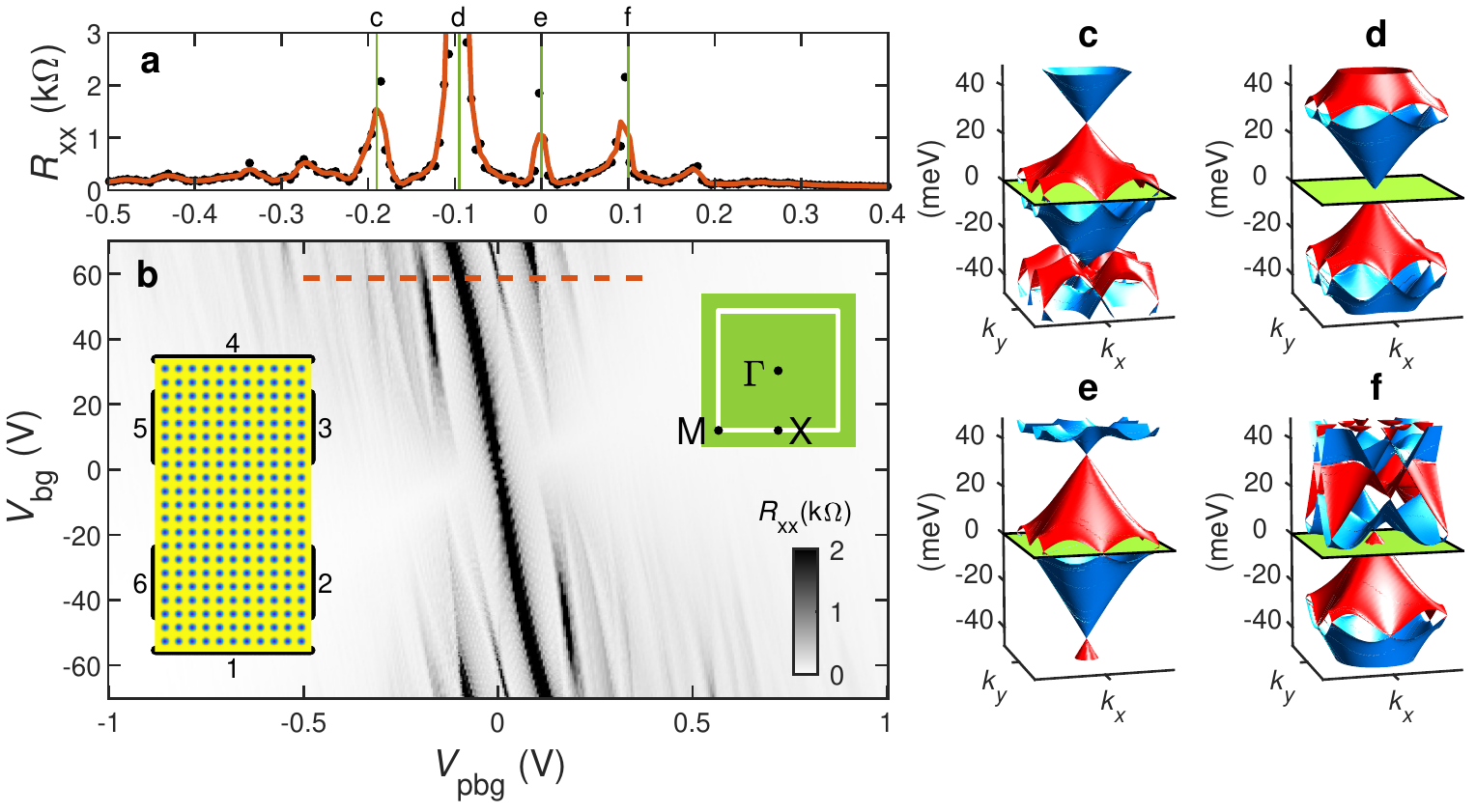}
\caption{\textbf{Transport Simulation and Band Structure.} \textbf{a} Four-terminal longitudinal resistance $R_{xx}$ taken at $V_\mathrm{bg} = 58$ V (dashed line in \textbf{b}). Continuous line shows smoothed data, and the black dots are the raw simulation results. Marked positions correspond to band structure plots. \textbf{b} $R_{xx}(V_\mathrm{pbg},V_\mathrm{bg})$ at zero magnetic field. Left inset: Geometry of the simulation. Right inset: Mini-Brillouin zone of the square superlattice with points of high symmetry. \textbf{c} to \textbf{f}: Band structure plots at positions marked in \textbf{a}. In each diagram, the green plane shows the position of the Fermi level.}
\label{fig:theozero}
\end{figure*}

To confirm these findings, we have performed quantum transport simulations (based on the Landauer-B\"{u}ttiker approach \cite{Datta1995}) and calculated miniband structures at $B=0$ T, both employing the realistic square superlattice potential obtained from finite-element-based electrostatic simulation taking into account the geometry of our device; see Supporting Information. Fig.~\ref{fig:theozero}a shows the simulated longitudinal resistance $R_{xx}$ as a function of $V_\mathrm{pbg}$ at fixed $V_\mathrm{bg}=58$~V for a six-terminal Hall bar (left inset of Fig.~\ref{fig:theozero}b). Several $R_{xx}$ peaks can be observed, around each of which clear sign changes of the Hall resistance $R_{xy}$ at $B=200$~mT are visible (see Supporting Information). Specifically, two $R_{xx}$ satellite Dirac peaks to the right of the main Dirac point and one to the left can clearly be seen in Fig.~\ref{fig:theozero}a, consistent with to the experimental data taken at large positive back gate voltage (Fig.~\ref{fig:ExpZero}e). This order reverses when tuning $V_\mathrm{bg}$ to negative, as seen in the simulated $R_{xx}$ map as a function of $V_\mathrm{pbg}$ and $V_\mathrm{bg}$ shown in Fig.~\ref{fig:theozero}b. Similar to Fig.~\ref{fig:ExpZero}d from the experiment, the resistance map of Fig.~\ref{fig:theozero}b shows a strong main Dirac peak throughout all back gate voltages, accompanied by a few faint lines parallel to the main Dirac line emerging with increasing $|V_\mathrm{bg}|$. A comparison of Fig.~\ref{fig:theozero}b with Fig.~\ref{fig:ExpZero}d shows an excellent agreement between our transport experiment and simulation.

The origins of the resistance peaks shown in Figs.~\ref{fig:theozero}a and b can be identified by comparing with our miniband structure calculations. A few examples are shown in Fig.~\ref{fig:theozero}c--f, where we have fixed the Fermi energy at zero represented by the green plane. The main Dirac peak is observed when the Fermi level lies exactly at the Dirac point of the unperturbed graphene band structure (d). The first satellite peak (c for the hole and e for the electron side) corresponds to well-resolved secondary and tertiary Dirac points at the X and M points of the (square) mini-Brillouin zone (right inset on Fig.~\ref{fig:theozero}b), while the second satellite resistance peak (f) is found when a large Dirac cone reappears at the center of the mini-Brillouin zone. It is somewhat obscured by higher bands intersecting the Fermi level. We provide a full movie of the miniband structures as $V_\mathrm{pbg}$ is swept through the line trace of the resistance in the Supporting Information. In addition to the features highlighted in Fig.~\ref{fig:theozero}, the movie shows that additional peaks are progressively harder to resolve as many minibands of electron and hole character are present at the Fermi level.

\begin{figure*}
\includegraphics[width=1\linewidth]{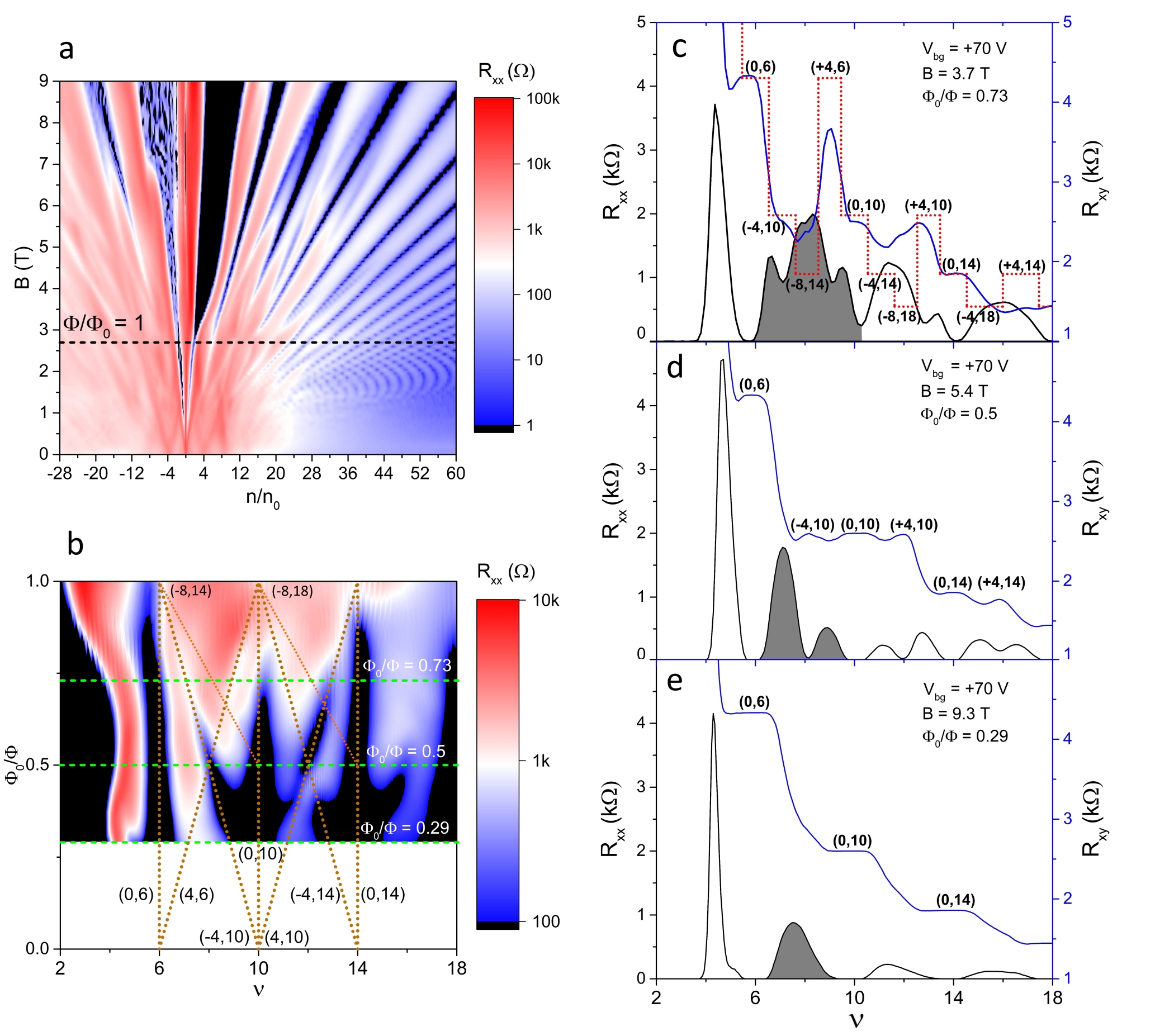}
\caption{\textbf{Experimental Magnetotransport Data at Large Fields.} \textbf{a} $R_{xx}$ plotted as a function of magnetic field and normalised charge carrier density $n/n_{0}$. Landau fans emerge from the main Dirac point and satellite Dirac points giving rise to superlattice induced energy gaps manifesting as additional minima in longitudinal resistance. The data were taken at $T=1.5$ K and a back gate voltage of $V_\mathrm{bg}=70$~V. \textbf{b} Replot of the data in \textbf{a} in a Wannier diagram with Landau level filling factor $\nu$ and inverse magnetic flux $\phi_{0}/\phi$. Vertical minima correspond to energy gaps between Landau levels, additional diagonal features are signatures of the largest energy gaps of the Hofstadter butterfly energy spectrum. Dotted lines highlight the Landau gaps (vertical) and the most pronounced Hofstadter gaps (diagonal) labeled by (s,t) in the second and third Landau level.	
\textbf{c} to \textbf{e} show linecuts at certain magnetic fields from the data in \textbf{b} showing the evolution of the magnetic bandstructure upon changing the magnetic flux per superlattice unit cell area. Energy gaps in the spectrum correspond to minima in $R_{xx}$ and plateaus in $R_{xy}$ and follow the equation $\nu=(\phi_{0}/\phi)s+t$. The observed energy gaps are labeled by their parameters $(s,t)$. Features with $s\neq0$ correspond to superlattice induced energy gaps. The dotted red line in \textbf{c} shows the ideal non-monotonic quantum Hall sequence evaluated for $s=0,\pm4$ and two additional energy gaps with $s=-8$.}
\label{fig:ExpHighB}
\end{figure*}

Now we focus on the high-field regime, where the Hofstadter energy spectrum is expected. In order to experimentally observe signatures of this energy spectrum the magnetic flux $\phi$ per superlattice unit cell area has to be of the order of one magnetic flux quantum $\phi_{0}$. For the studied sample with a square superlattice of nominal lattice constant $a=40$ nm, $\phi/\phi_{0}=1$ is reached experimentally at a magnetic field of $B=2.7$ T (corresponding to an actual lattice period of $a=39$ nm). Figure~\ref{fig:ExpHighB}a shows the longitudinal resistance as a function of normalised charge carrier density $n/n_{0}$ and magnetic fields up to 9 T. The magnetotransport data show additional features besides the main Landau fans, which emerge from the newly created satellite Dirac peaks as shown in Fig.~\ref{fig:ExpHighB}a. At higher fields the main Landau fan exhibits additional minima in longitudinal resistance which cannot be traced back to the original Dirac point and correspond to superlattice induced energy gaps.

To resolve the Hofstadter features better, in Fig.~\ref{fig:ExpHighB}b we replot part of the data in a Wannier diagram, {\em i.e.} the inverse magnetic flux $\phi_{0}/\phi$ versus the filling factor $\nu$. In this diagram, features belonging to quantum Hall states of the unpatterned graphene ($s=0,\ \nu=2,\ 6,\ 10,\ \dots$) appear as vertical lines while the energy gaps of the Hofstadter spectrum ($s\ne 0$) form diagonal lines. The largest energy gaps ($s=\pm 4$) are clearly resolved, following Eq. (\ref{Eq:Dio}).

\begin{figure*}
\includegraphics[width=1\linewidth]{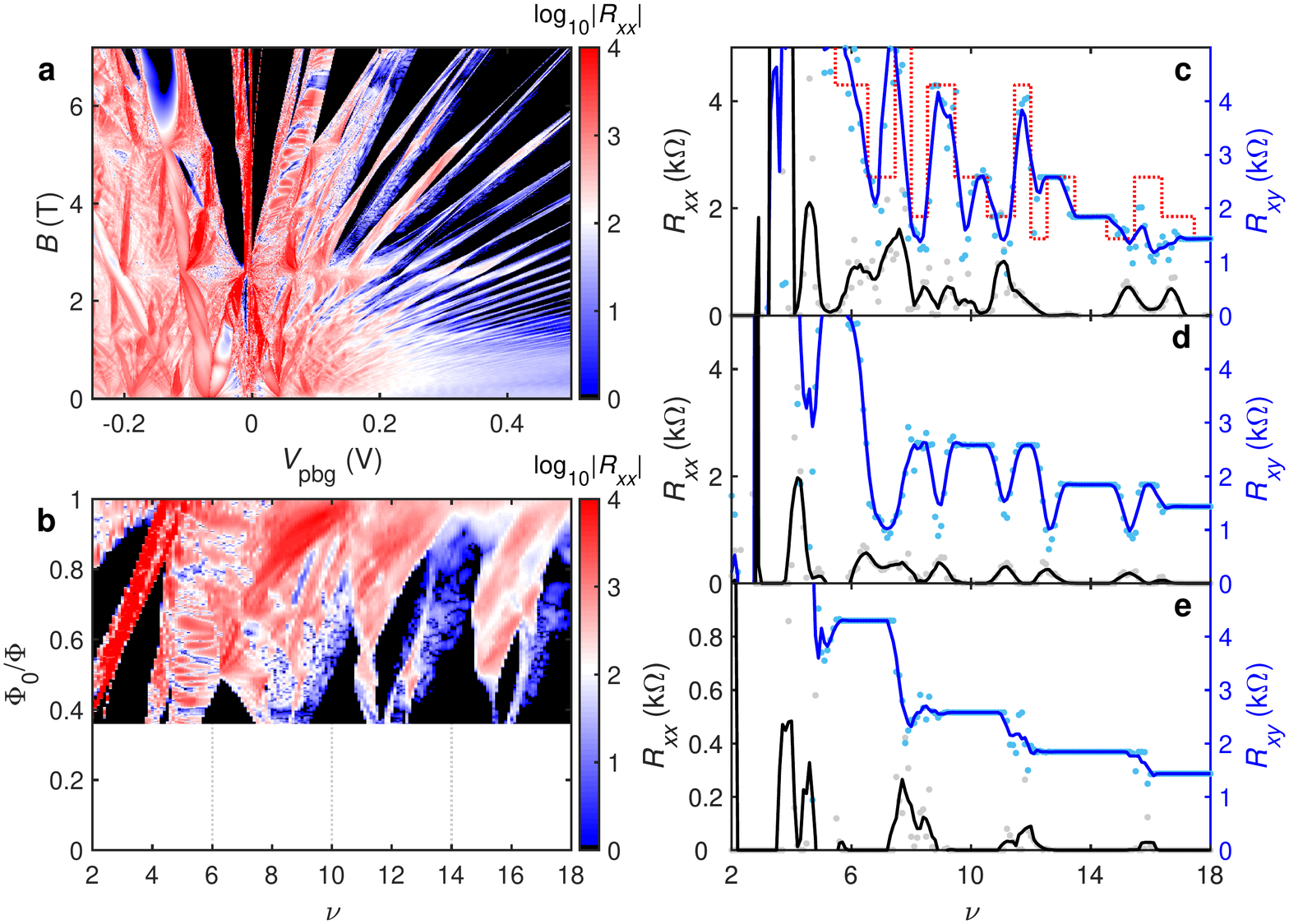}
\caption{\textbf{Simulated Transport at Large Fields.} Each panel corresponds to a similar panel in Fig.~\ref{fig:ExpHighB}. \textbf{a} Landau fan plot of $R_{xx}$ on a log scale at a fixed back gate voltage of $V_\mathrm{bg} = 50$ V. \textbf{b} Wannier diagram of $R_{xx}$. \textbf{c}-\textbf{e} Line cuts of the Hall (blue) and longitudinal resistance (black) at $\phi_0/\phi=0.73$, 0.50, and 0.36. Lines were smoothed over 5 neighbouring data points, raw data are shown as round dots. In panel \textbf{c}, the dotted red line shows the predicted positions of the quantum Hall plateaus, evaluated for $s=0,\pm4,\pm8$.}
\label{fig:theolarge}
\end{figure*}

The evolution of the magnetic band structure as a function of magnetic field becomes apparent by looking at the longitudinal and transverse resistance at certain values of magnetic flux per superlattice unit cell area (dashed horizontal lines in Fig.~\ref{fig:ExpHighB}b). At high magnetic fields ($\phi_{0}/\phi=0.29$), the usual sequence of quantum Hall plateaus can be observed, as expected for intrinsic monolayer graphene (see Fig.~\ref{fig:ExpHighB}e). By changing the magnetic flux ratio to $\phi_{0}/\phi=0.5$ a splitting of Landau bands into two minibands occurs. Additional minima in $R_{xx}$ can be observed accompanied by well developed quantum Hall plateaus (see Fig.~\ref{fig:ExpHighB}d). At a magnetic flux ratio of $\phi_{0}/\phi=0.73$ the quantum Hall resistance shows signatures of the two largest gaps ($s=\pm4$) of the Hofstadter butterfly (see Fig.~\ref{fig:ExpHighB}c). This is best visible in the second Landau band between $\nu=6$ and $\nu=10$ where $R_{xx}$ exhibits two additional minima, and the corresponding quantum Hall resistance $R_{xy}$ exhibits a non-trivial, non-monotonic sequence following the expected sequence of quantum Hall values if the Fermi energy is shifted into the largest energy gaps in the Hofstadter spectrum. \cite{Thouless1982} 
In addition there is also a tendency to approach the plateau value associated with $s=-8$, {\em i.e.} the next iteration of the self-similar spectrum. Also by looking at the second derivative of $R_{xx}$ with respect to $\nu$ traces corresponding to this energy gap become better visible and can be identified (see Supporting Information). These findings confirm the prediction by Thouless {\em et al.}\cite{Thouless1982} in a gate-defined superlattice in graphene.
 	
The situation at high magnetic fields was also modeled in transport calculations. Since the superlattice effect of our system comes solely from the electrostatic gating that results in a spatially periodic onsite energy profile, no symmetry-breaking terms are included in our tight-binding model Hamiltonian. Also, Zeeman interactions will be neglected in the present model calculations and, consequently, a possible lifting of graphene’s intrinsic 4-fold degeneracy is not taken into account, in accordance with the experimental results which show no apparent signatures of lifted degeneracies. Due to the restriction of the implemented scaled tight-binding approach\cite{Liu2015} (see Supporting Information), we consider a magnetic field range of up to $B=7.2$ T, which corresponds to $\phi_{0}/\phi=0.36$. For determining the electrostatic potential, we excluded the carrier density correction due to quantum capacitance, as there is no closed-form expression due to the graphene density of states deformed by finite magnetic field, contrary to the case at $B=0$ T. Therefore, the capacitance of both gates only contains the geometric capacitance resulting in a larger gate coupling. Nevertheless, by calculating the positions where the superlattice unit cells are filled with a multiple of four electrons or holes, we can match the features in the calculated Landau fan diagram with the experimental ones. This is done in Fig.~\ref{fig:theolarge}, where each panel of the experimental Fig.~\ref{fig:ExpHighB} matches a corresponding panel in the theory Fig.~\ref{fig:theolarge}. All data were calculated for $V_\mathrm{bg} = 50$ V. The similarity is quite striking. The Landau fan diagram (Fig.~\ref{fig:theolarge}a) shows Landau levels emanating from the main and the satellite Dirac peaks. For negative $V_\mathrm{pbg}$ (bipolar regime, opposite to $V_{bg}$), resistance values are higher, while on the positive $V_\mathrm{pbg}$ side, overall resistances are lower, and the energy gaps are better resolved. Additional simulation data at $V_\mathrm{bg} = 70$ V are shown in the Supporting Information.
To obtain the Wannier diagram in Fig.~\ref{fig:theolarge}b, we calculated the filling factor by taking straight lines from the main Dirac peak at $V_\mathrm{pbg} = -0.065$ V to the largest energy gaps at high field and density and interpolated from the raw data. 		
We note that close to the main Dirac peak, the Landau fan does not form straight lines, but starts out curved. The reason for this is unknown, but we speculate that in this region the carrier density in each unit cell is strongly non-uniform, leading to a deviation from the standard definition of the filling factor. Therefore, the disentangling of Landau and Hofstadter gaps for the calculated data is not as clear-cut as for the experimental data. Still we observe dark regions (low resistance) both for standard filling factors $\nu=2,\ 6,\ 10,\ 14,\ 18$ and additional Hofstadter gaps. 
When taking line traces of the Hall resistance at different flux ratios, the calculations reproduce the non-monotonic behavior of $R_{xy}$. For all flux ratios shown, we find a strong correspondence between experimental and numerical data. The peak splitting in $R_{xx}$ is also reproduced qualitatively,
but since the calculations were done without disorder, one cannot expect a one-to-one match.

In conclusion, we have presented an experimental realization of gate-tunable 2D superlattices in graphene with an arbitrary geometry, showing transport signatures of up to quaternary Dirac points, the Hofstadter energy spectrum and the non-trivial, non-monotonic quantum Hall response, all owing to the modified band structure.
Transport calculations at $B=0$ T and $B\ne 0$ T, based on a real-space Green's function method can reproduce the experimental data to high precision and support our interpretation. At $B=0$ T, the continuum model considering the same superlattice potential allows us to identify strong resistance features with isolated Dirac cones in the miniband structure, and explains why no further resistance peaks are observed. For large fields we calculate Landau fans corresponding to additional Dirac points, and also reproduce the non-monotonic quantum Hall response in detail.
The excellent correspondence between theory and experiment is striking and allows for a complete understanding of the observed transport features. Our experimental approach will also be suitable for studying more advanced nano-geometries, such as the Lieb lattice\cite{Lieb1989} or symmetric one-dimensional superlattices\cite{Kang2019}, which are not accessible by other methods.

\begin{acknowledgement}
M.-H.L.\ and S.-C.C.\ are supported by Taiwan Ministry of Science (107-2112-M-006-004-MY3 and 107-2627-E-006-001) and Ministry of Education (Higher Education Sprout Project).
Growth of hexagonal boron nitride crystals was supported by the
MEXT Element Strategy Initiative to Form Core Research Center, Grant
Number JPMXP0112101001 and the CREST(JPMJCR15F3), JST.
This  work  was  supported by the Deutsche Forschungsgemeinschaft (DFG, German Research Foundation) – project-id 314695032 – SFB 1277 (subprojects A07, A09) and GRK 1570.
\end{acknowledgement}

\textbf{Author Contributions}
R.H., D.W. and J.E. conceived the experiment, R.H., M.D. and A.S. developed the patterned gate preparation, R.H. fabricated the samples under study and performed transport measurements, K.W. and T.T. grew boron nitride crystals, M.-H.L. and K.R. developed the transport theory, M.-H.L. did transport calculations and S.-C.C. performed miniband calculations. R.H., J.E. and M.-H.L. wrote the manuscript with input from all co-authors.

\begin{suppinfo}
Sample fabrication and data acquisition, Details of the transport simulations and electronic structure calculations, Calculated gate maps and potential profiles, Additional simulated and experimental data, Movie of miniband structures
\end{suppinfo}


\begin{mcitethebibliography}{36}
	\providecommand*\natexlab[1]{#1}
	\providecommand*\mciteSetBstSublistMode[1]{}
	\providecommand*\mciteSetBstMaxWidthForm[2]{}
	\providecommand*\mciteBstWouldAddEndPuncttrue
	{\def\EndOfBibitem{\unskip.}}
	\providecommand*\mciteBstWouldAddEndPunctfalse
	{\let\EndOfBibitem\relax}
	\providecommand*\mciteSetBstMidEndSepPunct[3]{}
	\providecommand*\mciteSetBstSublistLabelBeginEnd[3]{}
	\providecommand*\EndOfBibitem{}
	\mciteSetBstSublistMode{f}
	\mciteSetBstMaxWidthForm{subitem}{(\alph{mcitesubitemcount})}
	\mciteSetBstSublistLabelBeginEnd
	{\mcitemaxwidthsubitemform\space}
	{\relax}
	{\relax}
	
	\bibitem[Keldysh(1963)]{Keldysh1963}
	Keldysh,~L.~V. Effect of ultrasonics on the electron spectrum of crystals.
	\emph{Soviet Physics Solid State} \textbf{1963}, \emph{4}, 1658\relax
	\mciteBstWouldAddEndPuncttrue
	\mciteSetBstMidEndSepPunct{\mcitedefaultmidpunct}
	{\mcitedefaultendpunct}{\mcitedefaultseppunct}\relax
	\EndOfBibitem
	\bibitem[Esaki and Tsu(1970)Esaki, and Tsu]{EsakiTsu1970}
	Esaki,~L.; Tsu,~R. Superlattice and Negative Differential Conductivity in
	Semiconductors. \emph{IBM Journal of Research and Development} \textbf{1970},
	\emph{14}, 61--65\relax
	\mciteBstWouldAddEndPuncttrue
	\mciteSetBstMidEndSepPunct{\mcitedefaultmidpunct}
	{\mcitedefaultendpunct}{\mcitedefaultseppunct}\relax
	\EndOfBibitem
	\bibitem[Hofstadter(1976)]{Hofstadter1976}
	Hofstadter,~D.~R. Energy levels and wave functions of Bloch electrons in
	rational and irrational magnetic fields. \emph{Phys. Rev. B} \textbf{1976},
	\emph{14}, 2239--2249\relax
	\mciteBstWouldAddEndPuncttrue
	\mciteSetBstMidEndSepPunct{\mcitedefaultmidpunct}
	{\mcitedefaultendpunct}{\mcitedefaultseppunct}\relax
	\EndOfBibitem
	\bibitem[MacDonald(1983)]{MacDonald1983}
	MacDonald,~A.~H. Landau-level subband structure of electrons on a square
	lattice. \emph{Phys. Rev. B} \textbf{1983}, \emph{28}, 6713--6717\relax
	\mciteBstWouldAddEndPuncttrue
	\mciteSetBstMidEndSepPunct{\mcitedefaultmidpunct}
	{\mcitedefaultendpunct}{\mcitedefaultseppunct}\relax
	\EndOfBibitem
	\bibitem[Thouless \latin{et~al.}(1982)Thouless, Kohmoto, Nightingale, and den
	Nijs]{Thouless1982}
	Thouless,~D.~J.; Kohmoto,~M.; Nightingale,~M.~P.; den Nijs,~M. Quantized Hall
	Conductance in a Two-Dimensional Periodic Potential. \emph{Phys. Rev. Lett.}
	\textbf{1982}, \emph{49}, 405--408\relax
	\mciteBstWouldAddEndPuncttrue
	\mciteSetBstMidEndSepPunct{\mcitedefaultmidpunct}
	{\mcitedefaultendpunct}{\mcitedefaultseppunct}\relax
	\EndOfBibitem
	\bibitem[Niu \latin{et~al.}(1985)Niu, Thouless, and Wu]{Niu1985}
	Niu,~Q.; Thouless,~D.~J.; Wu,~Y.-S. Quantized Hall conductance as a topological
	invariant. \emph{Phys. Rev. B} \textbf{1985}, \emph{31}, 3372--3377\relax
	\mciteBstWouldAddEndPuncttrue
	\mciteSetBstMidEndSepPunct{\mcitedefaultmidpunct}
	{\mcitedefaultendpunct}{\mcitedefaultseppunct}\relax
	\EndOfBibitem
	\bibitem[Aoki and Ando(1986)Aoki, and Ando]{Aoki1986}
	Aoki,~H.; Ando,~T. Universality of Quantum Hall Effect: Topological Invariant
	and Observable. \emph{Phys. Rev. Lett.} \textbf{1986}, \emph{57},
	3093--3096\relax
	\mciteBstWouldAddEndPuncttrue
	\mciteSetBstMidEndSepPunct{\mcitedefaultmidpunct}
	{\mcitedefaultendpunct}{\mcitedefaultseppunct}\relax
	\EndOfBibitem
	\bibitem[Wannier(1978)]{Wannier1978}
	Wannier,~G.~H. A Result Not Dependent on Rationality for Bloch Electrons in a
	Magnetic Field. \emph{physica status solidi (b)} \textbf{1978}, \emph{88},
	757--765\relax
	\mciteBstWouldAddEndPuncttrue
	\mciteSetBstMidEndSepPunct{\mcitedefaultmidpunct}
	{\mcitedefaultendpunct}{\mcitedefaultseppunct}\relax
	\EndOfBibitem
	\bibitem[Albrecht \latin{et~al.}(2001)Albrecht, Smet, von Klitzing, Weiss,
	Umansky, and Schweizer]{Albrecht2001}
	Albrecht,~C.; Smet,~J.~H.; von Klitzing,~K.; Weiss,~D.; Umansky,~V.;
	Schweizer,~H. Evidence of Hofstadter's Fractal Energy Spectrum in the
	Quantized Hall Conductance. \emph{Phys. Rev. Lett.} \textbf{2001}, \emph{86},
	147--150\relax
	\mciteBstWouldAddEndPuncttrue
	\mciteSetBstMidEndSepPunct{\mcitedefaultmidpunct}
	{\mcitedefaultendpunct}{\mcitedefaultseppunct}\relax
	\EndOfBibitem
	\bibitem[Geisler \latin{et~al.}(2004)Geisler, Smet, Umansky, von Klitzing,
	Naundorf, Ketzmerick, and Schweizer]{Geisler2004}
	Geisler,~M.~C.; Smet,~J.~H.; Umansky,~V.; von Klitzing,~K.; Naundorf,~B.;
	Ketzmerick,~R.; Schweizer,~H. Detection of a Landau Band-Coupling-Induced
	Rearrangement of the Hofstadter Butterfly. \emph{Phys. Rev. Lett.}
	\textbf{2004}, \emph{92}, 256801\relax
	\mciteBstWouldAddEndPuncttrue
	\mciteSetBstMidEndSepPunct{\mcitedefaultmidpunct}
	{\mcitedefaultendpunct}{\mcitedefaultseppunct}\relax
	\EndOfBibitem
	\bibitem[Dean \latin{et~al.}(2010)Dean, Young, Meric, Lee, Wang, Sorgenfrei,
	Watanabe, Taniguchi, Kim, Shepard, and Hone]{Dean2010}
	Dean,~C.~R.; Young,~A.~F.; Meric,~I.; Lee,~C.; Wang,~L.; Sorgenfrei,~S.;
	Watanabe,~K.; Taniguchi,~T.; Kim,~P.; Shepard,~K.~L.; Hone,~J. Boron nitride
	substrates for high-quality graphene electronics. \emph{Nature
		Nanotechnology} \textbf{2010}, \emph{5}, 722--\relax
	\mciteBstWouldAddEndPuncttrue
	\mciteSetBstMidEndSepPunct{\mcitedefaultmidpunct}
	{\mcitedefaultendpunct}{\mcitedefaultseppunct}\relax
	\EndOfBibitem
	\bibitem[Decker \latin{et~al.}(2011)Decker, Wang, Brar, Regan, Tsai, Wu,
	Gannett, Zettl, and Crommie]{Decker2011}
	Decker,~R.; Wang,~Y.; Brar,~V.~W.; Regan,~W.; Tsai,~H.-Z.; Wu,~Q.; Gannett,~W.;
	Zettl,~A.; Crommie,~M.~F. Local Electronic Properties of Graphene on a BN
	Substrate via Scanning Tunneling Microscopy. \emph{Nano Letters}
	\textbf{2011}, \emph{11}, 2291--2295\relax
	\mciteBstWouldAddEndPuncttrue
	\mciteSetBstMidEndSepPunct{\mcitedefaultmidpunct}
	{\mcitedefaultendpunct}{\mcitedefaultseppunct}\relax
	\EndOfBibitem
	\bibitem[Yankowitz \latin{et~al.}(2012)Yankowitz, Xue, Cormode,
	Sanchez-Yamagishi, Watanabe, Taniguchi, Jarillo-Herrero, Jacquod, and
	LeRoy]{Yankowitz2012}
	Yankowitz,~M.; Xue,~J.; Cormode,~D.; Sanchez-Yamagishi,~J.~D.; Watanabe,~K.;
	Taniguchi,~T.; Jarillo-Herrero,~P.; Jacquod,~P.; LeRoy,~B.~J. Emergence of
	superlattice Dirac points in graphene on hexagonal boron nitride.
	\emph{Nature Physics} \textbf{2012}, \emph{8}, 382--\relax
	\mciteBstWouldAddEndPuncttrue
	\mciteSetBstMidEndSepPunct{\mcitedefaultmidpunct}
	{\mcitedefaultendpunct}{\mcitedefaultseppunct}\relax
	\EndOfBibitem
	\bibitem[Wang \latin{et~al.}(2013)Wang, Meric, Huang, Gao, Gao, Tran,
	Taniguchi, Watanabe, Campos, Muller, Guo, Kim, Hone, Shepard, and
	Dean]{Wang1Dcontacts2013}
	Wang,~L.; Meric,~I.; Huang,~P.~Y.; Gao,~Q.; Gao,~Y.; Tran,~H.; Taniguchi,~T.;
	Watanabe,~K.; Campos,~L.~M.; Muller,~D.~A.; Guo,~J.; Kim,~P.; Hone,~J.;
	Shepard,~K.~L.; Dean,~C.~R. One-Dimensional Electrical Contact to a
	Two-Dimensional Material. \emph{Science} \textbf{2013}, \emph{342},
	614--617\relax
	\mciteBstWouldAddEndPuncttrue
	\mciteSetBstMidEndSepPunct{\mcitedefaultmidpunct}
	{\mcitedefaultendpunct}{\mcitedefaultseppunct}\relax
	\EndOfBibitem
	\bibitem[Dean \latin{et~al.}(2013)Dean, Wang, Maher, Forsythe, Ghahari, Gao,
	Katoch, Ishigami, Moon, Koshino, Taniguchi, Watanabe, Shepard, Hone, and
	Kim]{Dean2013}
	Dean,~C.~R.; Wang,~L.; Maher,~P.; Forsythe,~C.; Ghahari,~F.; Gao,~Y.;
	Katoch,~J.; Ishigami,~M.; Moon,~P.; Koshino,~M.; Taniguchi,~T.; Watanabe,~K.;
	Shepard,~K.~L.; Hone,~J.; Kim,~P. Hofstadter's butterfly and the fractal
	quantum Hall effect in moire superlattices. \emph{Nature} \textbf{2013},
	\emph{497}, 598--602\relax
	\mciteBstWouldAddEndPuncttrue
	\mciteSetBstMidEndSepPunct{\mcitedefaultmidpunct}
	{\mcitedefaultendpunct}{\mcitedefaultseppunct}\relax
	\EndOfBibitem
	\bibitem[Hunt \latin{et~al.}(2013)Hunt, Sanchez-Yamagishi, Young, Yankowitz,
	LeRoy, Watanabe, Taniguchi, Moon, Koshino, Jarillo-Herrero, and
	Ashoori]{Hunt2013}
	Hunt,~B.; Sanchez-Yamagishi,~J.~D.; Young,~A.~F.; Yankowitz,~M.; LeRoy,~B.~J.;
	Watanabe,~K.; Taniguchi,~T.; Moon,~P.; Koshino,~M.; Jarillo-Herrero,~P.;
	Ashoori,~R.~C. Massive Dirac Fermions and Hofstadter Butterfly in a van der
	Waals Heterostructure. \emph{Science} \textbf{2013}, \emph{340},
	1427--1430\relax
	\mciteBstWouldAddEndPuncttrue
	\mciteSetBstMidEndSepPunct{\mcitedefaultmidpunct}
	{\mcitedefaultendpunct}{\mcitedefaultseppunct}\relax
	\EndOfBibitem
	\bibitem[Ponomarenko \latin{et~al.}(2013)Ponomarenko, Gorbachev, Yu, Elias,
	Jalil, Patel, Mishchenko, Mayorov, Woods, Wallbank, Mucha-Kruczynski, Piot,
	Potemski, Grigorieva, Novoselov, Guinea, Fal'ko, and Geim]{Ponomarenko2013}
	Ponomarenko,~L.~A. \latin{et~al.}  Cloning of Dirac fermions in graphene
	superlattices. \emph{Nature} \textbf{2013}, \emph{497}, 594--597\relax
	\mciteBstWouldAddEndPuncttrue
	\mciteSetBstMidEndSepPunct{\mcitedefaultmidpunct}
	{\mcitedefaultendpunct}{\mcitedefaultseppunct}\relax
	\EndOfBibitem
	\bibitem[Wallbank \latin{et~al.}(2013)Wallbank, Patel,
	Mucha-Kruczy\ifmmode~\acute{n}\else \'{n}\fi{}ski, Geim, and
	Fal'ko]{Wallbank2013}
	Wallbank,~J.~R.; Patel,~A.~A.; Mucha-Kruczy\ifmmode~\acute{n}\else
	\'{n}\fi{}ski,~M.; Geim,~A.~K.; Fal'ko,~V.~I. Generic miniband structure of
	graphene on a hexagonal substrate. \emph{Phys. Rev. B} \textbf{2013},
	\emph{87}, 245408\relax
	\mciteBstWouldAddEndPuncttrue
	\mciteSetBstMidEndSepPunct{\mcitedefaultmidpunct}
	{\mcitedefaultendpunct}{\mcitedefaultseppunct}\relax
	\EndOfBibitem
	\bibitem[Moon and Koshino(2014)Moon, and Koshino]{Moon2014}
	Moon,~P.; Koshino,~M. Electronic properties of graphene/hexagonal-boron-nitride
	moir\'e superlattice. \emph{Phys. Rev. B} \textbf{2014}, \emph{90},
	155406\relax
	\mciteBstWouldAddEndPuncttrue
	\mciteSetBstMidEndSepPunct{\mcitedefaultmidpunct}
	{\mcitedefaultendpunct}{\mcitedefaultseppunct}\relax
	\EndOfBibitem
	\bibitem[Eroms and Weiss(2009)Eroms, and Weiss]{EromsNJP2009}
	Eroms,~J.; Weiss,~D. Weak localization and transport gap in graphene antidot
	lattices. \emph{New Journal of Physics} \textbf{2009}, \emph{11},
	095021\relax
	\mciteBstWouldAddEndPuncttrue
	\mciteSetBstMidEndSepPunct{\mcitedefaultmidpunct}
	{\mcitedefaultendpunct}{\mcitedefaultseppunct}\relax
	\EndOfBibitem
	\bibitem[Sandner \latin{et~al.}(2015)Sandner, Preis, Schell, Giudici, Watanabe,
	Taniguchi, Weiss, and Eroms]{Sandner2015}
	Sandner,~A.; Preis,~T.; Schell,~C.; Giudici,~P.; Watanabe,~K.; Taniguchi,~T.;
	Weiss,~D.; Eroms,~J. Ballistic Transport in Graphene Antidot Lattices.
	\emph{Nano Letters} \textbf{2015}, \emph{15}, 8402--8406\relax
	\mciteBstWouldAddEndPuncttrue
	\mciteSetBstMidEndSepPunct{\mcitedefaultmidpunct}
	{\mcitedefaultendpunct}{\mcitedefaultseppunct}\relax
	\EndOfBibitem
	\bibitem[Yagi \latin{et~al.}(2015)Yagi, Sakakibara, Ebisuoka, Onishi, Watanabe,
	Taniguchi, and Iye]{Yagi2015}
	Yagi,~R.; Sakakibara,~R.; Ebisuoka,~R.; Onishi,~J.; Watanabe,~K.;
	Taniguchi,~T.; Iye,~Y. Ballistic transport in graphene antidot lattices.
	\emph{Phys. Rev. B} \textbf{2015}, \emph{92}, 195406\relax
	\mciteBstWouldAddEndPuncttrue
	\mciteSetBstMidEndSepPunct{\mcitedefaultmidpunct}
	{\mcitedefaultendpunct}{\mcitedefaultseppunct}\relax
	\EndOfBibitem
	\bibitem[Power \latin{et~al.}(2017)Power, Thomsen, Jauho, and
	Pedersen]{Power2017}
	Power,~S.~R.; Thomsen,~M.~R.; Jauho,~A.-P.; Pedersen,~T.~G. Electron
	trajectories and magnetotransport in nanopatterned graphene under
	commensurability conditions. \emph{Phys. Rev. B} \textbf{2017}, \emph{96},
	075425\relax
	\mciteBstWouldAddEndPuncttrue
	\mciteSetBstMidEndSepPunct{\mcitedefaultmidpunct}
	{\mcitedefaultendpunct}{\mcitedefaultseppunct}\relax
	\EndOfBibitem
	\bibitem[Datseris \latin{et~al.}(2019)Datseris, Geisel, and
	Fleischmann]{Datseris2019}
	Datseris,~G.; Geisel,~T.; Fleischmann,~R. Robustness of ballistic transport in
	antidot superlattices. \emph{New Journal of Physics} \textbf{2019},
	\emph{21}, 043051\relax
	\mciteBstWouldAddEndPuncttrue
	\mciteSetBstMidEndSepPunct{\mcitedefaultmidpunct}
	{\mcitedefaultendpunct}{\mcitedefaultseppunct}\relax
	\EndOfBibitem
	\bibitem[Park \latin{et~al.}(2008)Park, Yang, Son, Cohen, and
	Louie]{ParkPeriodic2008}
	Park,~C.-H.; Yang,~L.; Son,~Y.-W.; Cohen,~M.~L.; Louie,~S.~G. New Generation of
	Massless Dirac Fermions in Graphene under External Periodic Potentials.
	\emph{Phys. Rev. Lett.} \textbf{2008}, \emph{101}, 126804\relax
	\mciteBstWouldAddEndPuncttrue
	\mciteSetBstMidEndSepPunct{\mcitedefaultmidpunct}
	{\mcitedefaultendpunct}{\mcitedefaultseppunct}\relax
	\EndOfBibitem
	\bibitem[Jessen \latin{et~al.}(2019)Jessen, Gammelgaard, Thomsen, Mackenzie,
	Thomsen, Caridad, Duegaard, Watanabe, Taniguchi, Booth, Pedersen, Jauho, and
	B{\o}ggild]{Jessen2019}
	Jessen,~B.~S.; Gammelgaard,~L.; Thomsen,~M.~R.; Mackenzie,~D. M.~A.;
	Thomsen,~J.~D.; Caridad,~J.~M.; Duegaard,~E.; Watanabe,~K.; Taniguchi,~T.;
	Booth,~T.~J.; Pedersen,~T.~G.; Jauho,~A.-P.; B{\o}ggild,~P. Lithographic band
	structure engineering of graphene. \emph{Nature Nanotechnology}
	\textbf{2019}, \emph{14}, 340--346\relax
	\mciteBstWouldAddEndPuncttrue
	\mciteSetBstMidEndSepPunct{\mcitedefaultmidpunct}
	{\mcitedefaultendpunct}{\mcitedefaultseppunct}\relax
	\EndOfBibitem
	\bibitem[Forsythe \latin{et~al.}(2018)Forsythe, Zhou, Watanabe, Taniguchi,
	Pasupathy, Moon, Koshino, Kim, and Dean]{Forsythe2018}
	Forsythe,~C.; Zhou,~X.; Watanabe,~K.; Taniguchi,~T.; Pasupathy,~A.; Moon,~P.;
	Koshino,~M.; Kim,~P.; Dean,~C.~R. Band structure engineering of 2D materials
	using patterned dielectric superlattices. \emph{Nature Nanotechnology}
	\textbf{2018}, \emph{13}, 566--571\relax
	\mciteBstWouldAddEndPuncttrue
	\mciteSetBstMidEndSepPunct{\mcitedefaultmidpunct}
	{\mcitedefaultendpunct}{\mcitedefaultseppunct}\relax
	\EndOfBibitem
	\bibitem[Forsythe(2017)]{ForsythePhD}
	Forsythe,~C. Fractal Hofstadter Band Structure in Patterned Dielectric
	Superlattice Graphene Systems. Ph.D.\ thesis, ColumbiaUniversity, 2017\relax
	\mciteBstWouldAddEndPuncttrue
	\mciteSetBstMidEndSepPunct{\mcitedefaultmidpunct}
	{\mcitedefaultendpunct}{\mcitedefaultseppunct}\relax
	\EndOfBibitem
	\bibitem[Drienovsky \latin{et~al.}(2017)Drienovsky, Sandner, Baumgartner, Liu,
	Taniguchi, Watanabe, Richter, Weiss, and Eroms]{DrienovskyPBG2017}
	Drienovsky,~M.; Sandner,~A.; Baumgartner,~C.; Liu,~M.-H.; Taniguchi,~T.;
	Watanabe,~K.; Richter,~K.; Weiss,~D.; Eroms,~J. arxiv:1703.05631\relax
	\mciteBstWouldAddEndPuncttrue
	\mciteSetBstMidEndSepPunct{\mcitedefaultmidpunct}
	{\mcitedefaultendpunct}{\mcitedefaultseppunct}\relax
	\EndOfBibitem
	\bibitem[Drienovsky \latin{et~al.}(2018)Drienovsky, Joachimsmeyer, Sandner,
	Liu, Taniguchi, Watanabe, Richter, Weiss, and Eroms]{Drienovsky2018}
	Drienovsky,~M.; Joachimsmeyer,~J.; Sandner,~A.; Liu,~M.-H.; Taniguchi,~T.;
	Watanabe,~K.; Richter,~K.; Weiss,~D.; Eroms,~J. Commensurability Oscillations
	in One-Dimensional Graphene Superlattices. \emph{Phys. Rev. Lett.}
	\textbf{2018}, \emph{121}, 026806\relax
	\mciteBstWouldAddEndPuncttrue
	\mciteSetBstMidEndSepPunct{\mcitedefaultmidpunct}
	{\mcitedefaultendpunct}{\mcitedefaultseppunct}\relax
	\EndOfBibitem
	\bibitem[Liu(2013)]{MingHaoCap2013}
	Liu,~M.-H. Theory of carrier density in multigated doped graphene sheets with
	quantum correction. \emph{Phys. Rev. B} \textbf{2013}, \emph{87},
	125427\relax
	\mciteBstWouldAddEndPuncttrue
	\mciteSetBstMidEndSepPunct{\mcitedefaultmidpunct}
	{\mcitedefaultendpunct}{\mcitedefaultseppunct}\relax
	\EndOfBibitem
	\bibitem[Datta(1995)]{Datta1995}
	Datta,~S. \emph{{Electronic Transport in Mesoscopic Systems}}; Cambridge
	University Press: Cambridge, 1995\relax
	\mciteBstWouldAddEndPuncttrue
	\mciteSetBstMidEndSepPunct{\mcitedefaultmidpunct}
	{\mcitedefaultendpunct}{\mcitedefaultseppunct}\relax
	\EndOfBibitem
	\bibitem[Liu \latin{et~al.}(2015)Liu, Rickhaus, Makk, T{\'o}v{\'a}ri, Maurand,
	Tkatschenko, Weiss, Sch{\"o}nenberger, and Richter]{Liu2015}
	Liu,~M.-H.; Rickhaus,~P.; Makk,~P.; T{\'o}v{\'a}ri,~E.; Maurand,~R.;
	Tkatschenko,~F.; Weiss,~M.; Sch{\"o}nenberger,~C.; Richter,~K. {Scalable
		Tight-Binding Model for Graphene}. \emph{Phys. Rev. Lett.} \textbf{2015},
	\emph{114}, 036601\relax
	\mciteBstWouldAddEndPuncttrue
	\mciteSetBstMidEndSepPunct{\mcitedefaultmidpunct}
	{\mcitedefaultendpunct}{\mcitedefaultseppunct}\relax
	\EndOfBibitem
	\bibitem[Lieb(1989)]{Lieb1989}
	Lieb,~E.~H. Two theorems on the Hubbard model. \emph{Phys. Rev. Lett.}
	\textbf{1989}, \emph{62}, 1201--1204\relax
	\mciteBstWouldAddEndPuncttrue
	\mciteSetBstMidEndSepPunct{\mcitedefaultmidpunct}
	{\mcitedefaultendpunct}{\mcitedefaultseppunct}\relax
	\EndOfBibitem
	\bibitem[Kang \latin{et~al.}(2019)Kang, Chen, and Liu]{Kang2019}
	Kang,~W.-H.; Chen,~S.-C.; Liu,~M.-H. arxiv:1912:13202\relax
	\mciteBstWouldAddEndPuncttrue
	\mciteSetBstMidEndSepPunct{\mcitedefaultmidpunct}
	{\mcitedefaultendpunct}{\mcitedefaultseppunct}\relax
	\EndOfBibitem
\end{mcitethebibliography}
\providecommand{\latin}[1]{#1}
\makeatletter
\providecommand{\doi}
{\begingroup\let\do\@makeother\dospecials
	\catcode`\{=1 \catcode`\}=2 \doi@aux}
\providecommand{\doi@aux}[1]{\endgroup\texttt{#1}}
\makeatother
\providecommand*\mcitethebibliography{\thebibliography}
\csname @ifundefined\endcsname{endmcitethebibliography}
{\let\endmcitethebibliography\endthebibliography}{}

\newpage

\end{document}